\documentclass{moriond}

\def\be{\begin{equation}}
\def\ee{\end{equation}}
\def\bea{\begin{eqnarray}}
\def\eea{\end{eqnarray}}

\newcommand{\Photo}{\includegraphics[height=35mm]{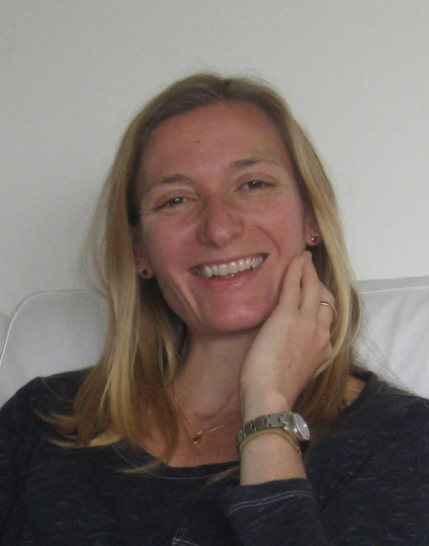}}
\begin{document}
\vspace*{4cm}
\title{QCD and High Energy Interactions: Moriond 2015 Theory Summary}

\author{ G.ZANDERIGHI }

\address{CERN, PH-TH, CH-1211 Geneva 23, Switzerland\\ and \\
Department of Physics, Theoretical Physics, 1 Keble Road,Oxford OX1 3NP, England}

\maketitle\abstracts{ I will summarise the new theory
  developments that emerged during the 2015 QCD Moriond conference. I
  will give my perspective on some of the topics and emphasise what I
  consider most relevant.  }

\section{Introduction}
We had more than 30 theory talks covering a rather broad range of
topics. The theory talks were allocated to different sessions: Higgs
and top, Flavour, QCD, the latter divided into more
perturbative/technical aspects and more non-perturbative/formal
developments, New Phenomena and Heavy Ions.  Sections in this summary
reflect the sessions we had at the conference.

\section{Higgs and top}

Remarkably, we had only one theory talk in the Higgs and top
section.~\footnote{Next-to-next-to-leading order results for top-pair production were presented
  in the QCD session.} Bernhard Mistlberger presented first N$^3$LO
results for the inclusive gluon-fusion Higgs cross section in the infinite top-mass effective theory.~\cite{mistlberger}  This
calculation is, in my opinion, the theory highlight of the meeting,
hence I will spend few words on this topic.

\begin{figure}
  \includegraphics[angle=0,width=0.48\linewidth]{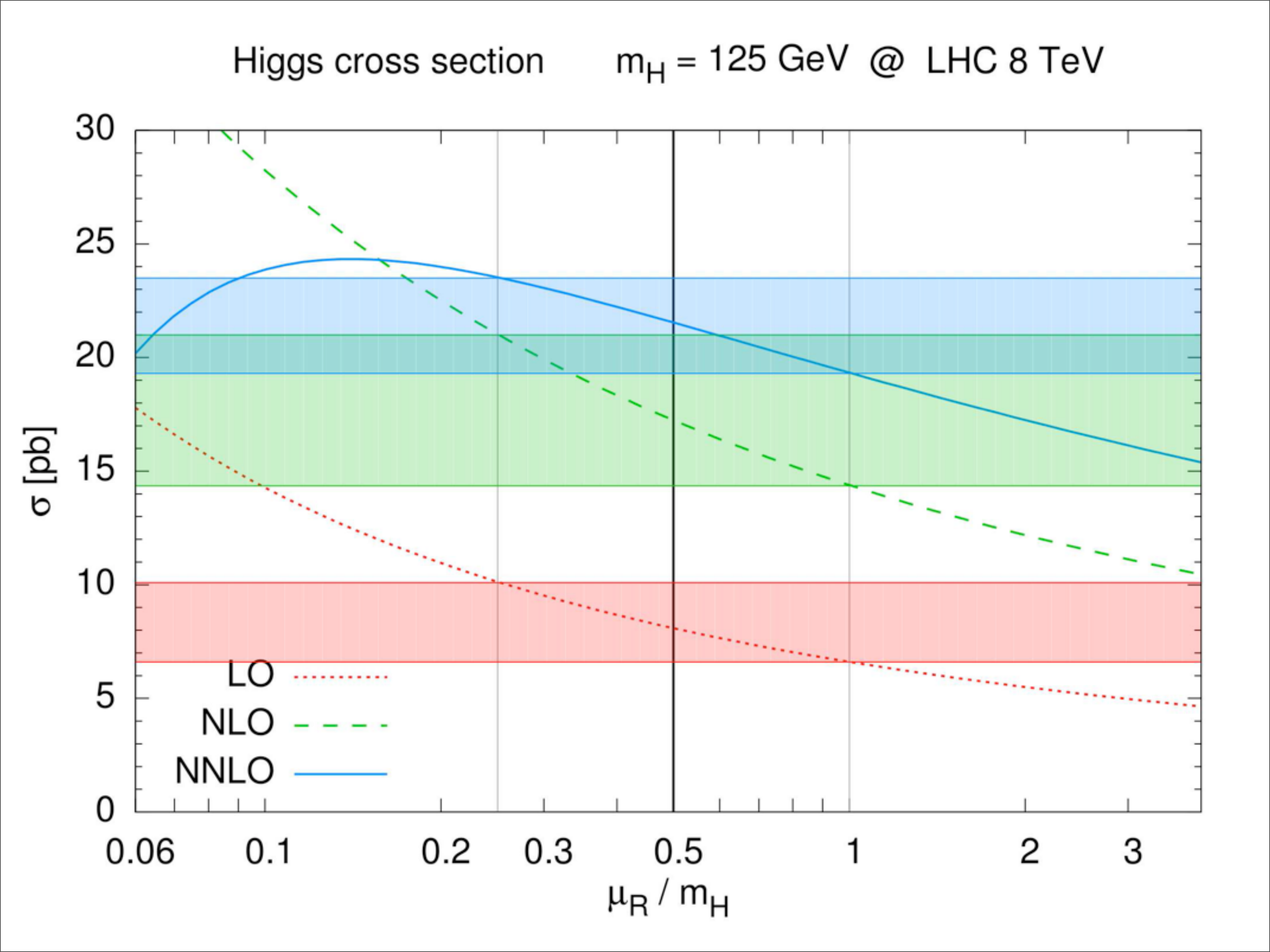}
\hfill
\includegraphics[angle=0,width=0.48\linewidth]{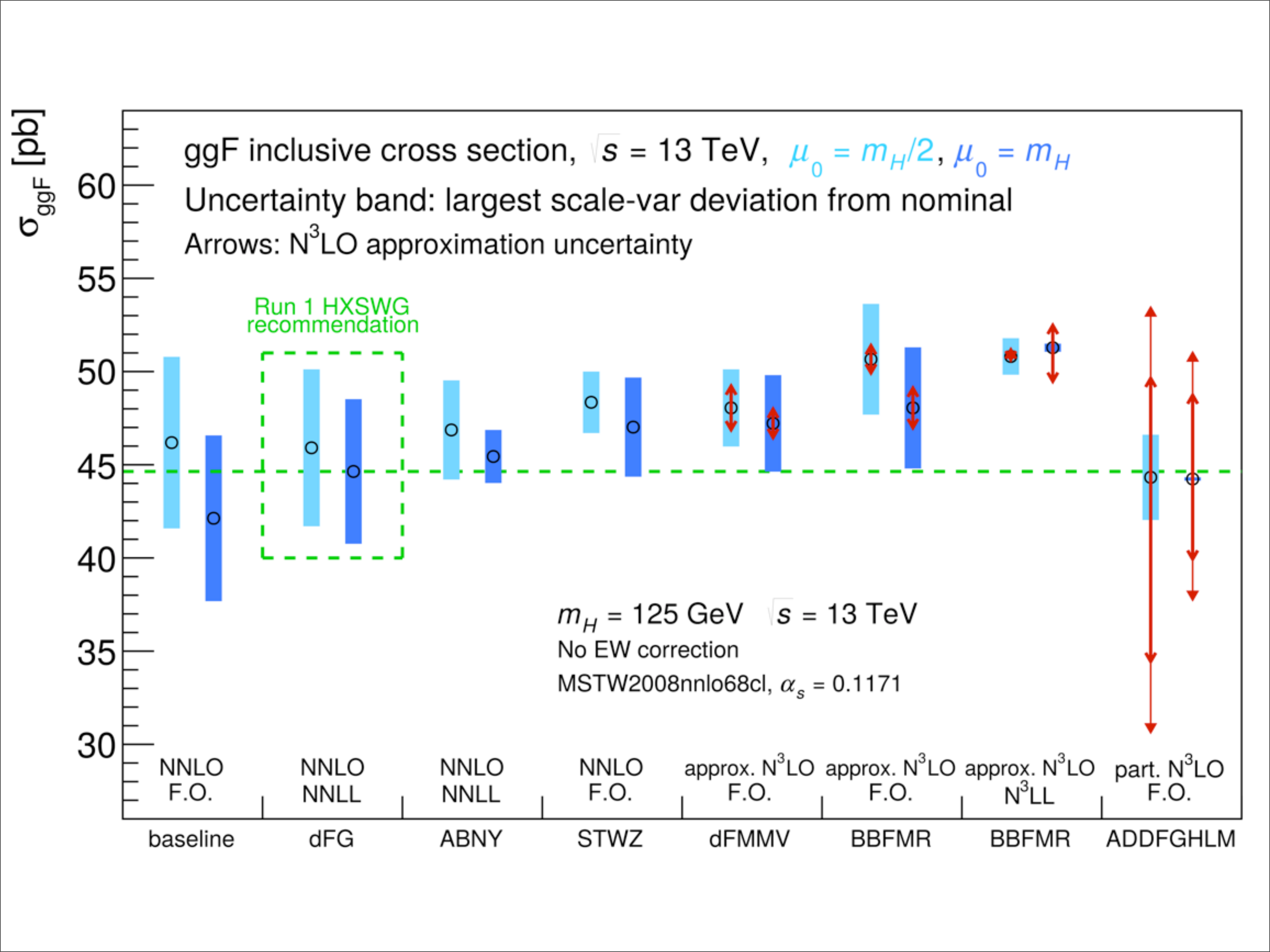}
\caption[]{Left plot: total gluon-fusion Higgs cross section at the
  LHC (8 TeV) as a function of the renormalisation scale at various
  orders in perturbation theory.  The plot has been obtained using the code of ref.~\cite{Ball:2013bra}.
  Right plot: a comparison of
  predictions for the total gluon-fusion Higgs cross section at the
  LHC (13 TeV) from various groups.  }\label{fig:sigtot}
\end{figure}

\begin{figure}[htb]
\includegraphics[angle=0,width=0.48\linewidth]{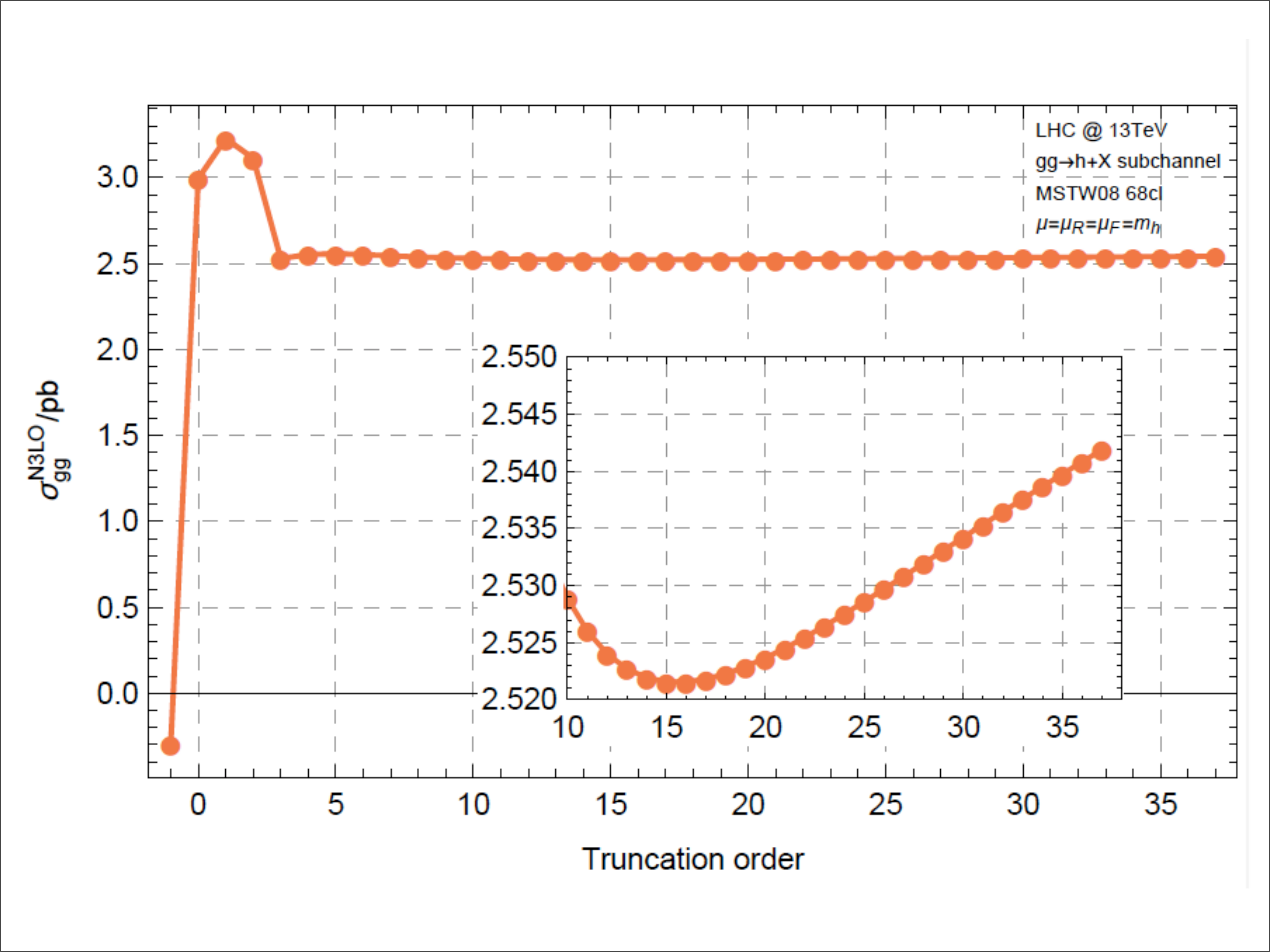}
\hfill
\includegraphics[angle=0,width=0.48\linewidth]{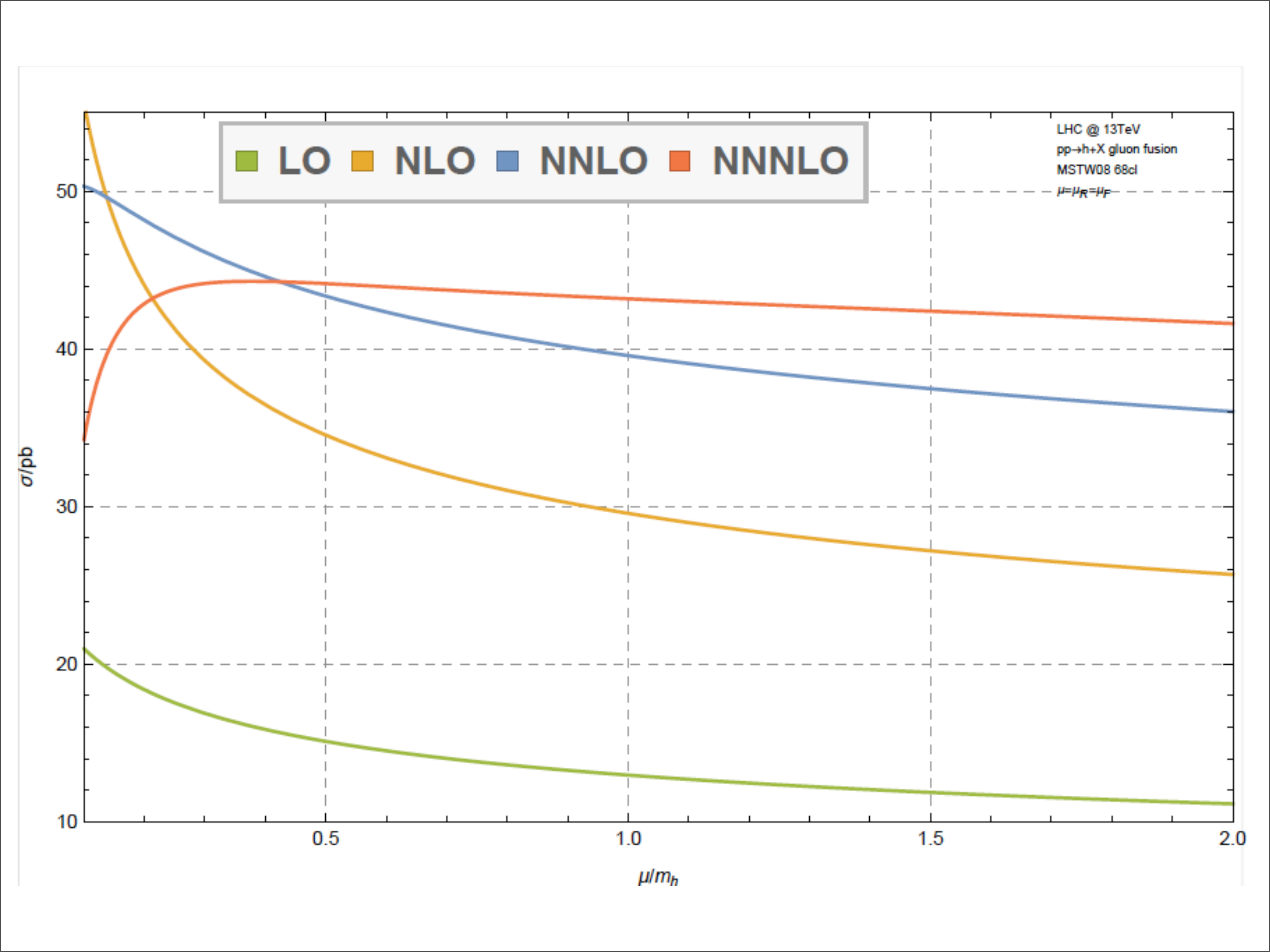}
\caption[]{Left plot: the N$^3$LO correction from the $gg$ channel to
  the total gluon-fusion Higgs cross section as a function of the
  number of terms included in the threshold expansion.  Right plot:
  scale variation for the gluon-fusion cross section at all
  perturbative orders through N$^3$LO.  }\label{fig:sigNNNLO}
\end{figure}

In order to put this work into context, it is useful to examine the
left panel of Fig.~\ref{fig:sigtot}, 
which shows the slow perturbative
convergence of the Higgs cross section. Furthermore, it is evident
from the figure that the renormalisation (and factorisation) scale
variation, that are commonly used to estimate theory uncertainties,
underestimate the shift between different perturbative
orders. Fig.~\ref{fig:sigtot} (right), presented at the general
assembly meeting of the Higgs cross section working group in
January,~\cite{higgsmeeting} shows results for the preferred total
gluon-fusion cross section from different groups. Each group provided
a prediction for the cross section obtained by using as a central
renormalisation and factorisation scale choice $m_H/2$ (light blue)
and $m_H$ (dark blue). The bands illustrate the scale uncertainty,
obtained by varying renormalisation and factorisation scales
independently by a factor 2 (avoiding the variation where they differ
by a factor 4), while the red errors denote the total uncertainty on
the numbers as estimated by the groups.  It is clear from the plot
that there was no consensus on the size of the uncertainty on this
cross section. This becomes particularly evident from the uncertainties quoted by 
the last two groups. However, the amount of perturbative control on
this cross section has a direct impact on a range of new physics
searches in the Higgs sector, hence it was crucial to improve on these
predictions by computing the cross section at N$^3$LO. This
calculation is however extremely challenging. In fact, the computation involves ${\cal O}(10^5)$ interference diagrams (for comparison only 1000 at NNLO), about 60 millions of loop
and phase space integrals (47000 at NNLO) and about 1000 master
integrals (26 at NNLO).  The calculation was performed as an expansion
around the threshold, where up to 37 terms in the expansion could be
computed. This result is shown in the left panel of
Fig.~\ref{fig:sigNNNLO}, 
while the right panel shows the dependence of
the cross section on the renormalisation and factorisation scales
(varied together) at all
  perturbative orders through N$^3$LO. The numbers to take home are that the N$^3$LO
corrections amount to about 2\% at scale $M_H/2$ and the residual
uncertainty as estimated from scale variation is also about 2-3\%. 
At this level of precision, other uncertainties (errors on parton
distribution functions, treatment of electroweak corrections, exact
top-mass corrections beyond the heavy-top approximation, top-bottom
interference in loops...) now become all important. 
Updated predictions, that  will also include an independent scale variation, will provide 
a more robust estimate of the uncertainty due to missing higher orders. 
Nonetheless, the very little experience that we have with scale variation at this order may suggest a conservative approach.

\section{Flavour}

The flavour day was possibly the most exciting day of the conference
because of the flavour anomalies observed recently at the LHC. 
Nazila Mahmoudi presented a concise introduction to flavour physics, recalling in particular the reasons why the  flavour physics is so rich and interesting.~\cite{mahmoudi} 
First of all, flavour physics is sensitive
to new physics (NP) energies scales that are much larger than the
collider energy, hence through flavour physics one could probe NP
before it is observed directly in collider experiments. Furthermore,
CP violation is closely related to flavour physics: the only source of
CP violation in the Standard Model (SM) comes from the Cabibbo-Kobayashi-Maskawa (CKM) matrix, but
for baryogenesis we know that we need other sources of CP
violation. On top of this, there is the  "SM flavour puzzle", i.e. the origin of masses and mixing of quarks and leptons, and the "new physics flavour puzzle", i.e. the mechanism protecting TeV-scale NP from causing large deviations from the SM predictions in the flavour observables that we have measured so far. 
Recently, there has been a lot of new
data in this sector mainly from LHCb, but also from ATLAS and CMS.
One of the new LHCb measurements concerns the CKM element $V_{ub}$. There has been a
longstanding tension in $V_{ub}$ (but also in $V_{cb}$) from inclusive
and exclusive decays. It was long believed that this measurement was
not possible at LHCb, yet $V_{ub}$ was recently measured there. This
measurement, which was presented for the first time at the Electroweak
Moriond meeting this year, seems to confirm the exclusive measurement.

Looking at other LHC data, almost all measurements are currently
consistent with the SM. Yet, recently four hints for new physics in
the flavour sector have been reported:
\begin{itemize}
\item the branching ratio $H\to \mu \tau$ was measured to be $ (0.84 \pm
  0.40)\%$ -  rather than 0;
\item in the decay $B \to K^* \mu^+\mu^-$ an anomaly was observed in
  an angular distribution called $P_5^\prime$;
\item the SM branching ratio $B_s \to \phi \mu^+\mu^-$ at high invariant mass 
  is above measurements;
\item the ratio $R_K\equiv {\rm Br}(B^+ \to K^+ \mu^+\mu^- )/{\rm Br}(B^+ \to K^+ e^+ e^- )$
  was measured to be $0.75 \pm 0.10$, rather than 1.
\end{itemize}
Adolfo Guevara focused on the anomaly in $R(K)$ and showed that the
discrepancy can not be attributed to long-distance, poorly-modelled
effects.~\cite{guevara} While this is not an exciting finding, it is
of course very important to have a solid estimate on long-distance
effects.
Ben Grinstein presented a very concise introduction to effective field
theories (EFTs), and stressed that the reason for using EFTs is that
they often have more predictive power.~\cite{grinstein} This is due
both, to the fact that they involve less free parameters and that you
can often simplify (technically challenging) calculations.~ In
particular, in terms of dimension 6 operators one can write the
following contributions to the Lagrangian:
\begin{equation}
{\cal L}{\rm eff} = -\frac{G_F}{\sqrt{2}} \sum_{p=u,c}\lambda_{ps}\left (C_1 {\cal O}_1^p+C_2 {\cal O}_2^p +\sum_{i=3}^{10} C_i  {\cal O}_i\right)\,. 
\end{equation}
For the above anomalies, of particular interest are the electromagnetic dipole, the vector
and axial-vector operators ${\cal O}_7$, ${\cal O}_9$ and ${\cal
  O}_{10}$, respectively
\begin{equation}
{\cal O}_7 = \frac{e}{(4\pi)^2} {m}_b \left[ \overline{s} \sigma^{\mu\nu} P_R b\right] F_{\mu\nu},\>\,\,
{\cal O}_9 = \frac{e^2}{(4\pi)^2}\left[ \overline{s} \gamma^{\mu} P_L b\right] \left[ \overline l \gamma^\mu l\right],\>\,\,
{\cal O}_{10} = \frac{e^2}{(4\pi)^2}\left[ \overline{s} \gamma^{\mu} P_L b\right] \left[ \overline l \gamma^\mu \gamma_5l\right]\,.
\end{equation}
Grinstein pointed out that taking into account all bounds, and assuming that
NP effects are only due to scalar and tensor semi-leptonic operators one can
constraint $0.982\leq R_K \le 1.007$. On the other hand, the measured
value of $R_K$ can be explained with a correction to the Wilson coefficient of the
vector operator, $C_9^{\rm NP}\approx -1$ (other explanations for the
measured value of $R_K$ could also come from a $Z'$ or leptoquarks).

\begin{figure}[t]
\includegraphics[angle=0,width=0.49\linewidth]{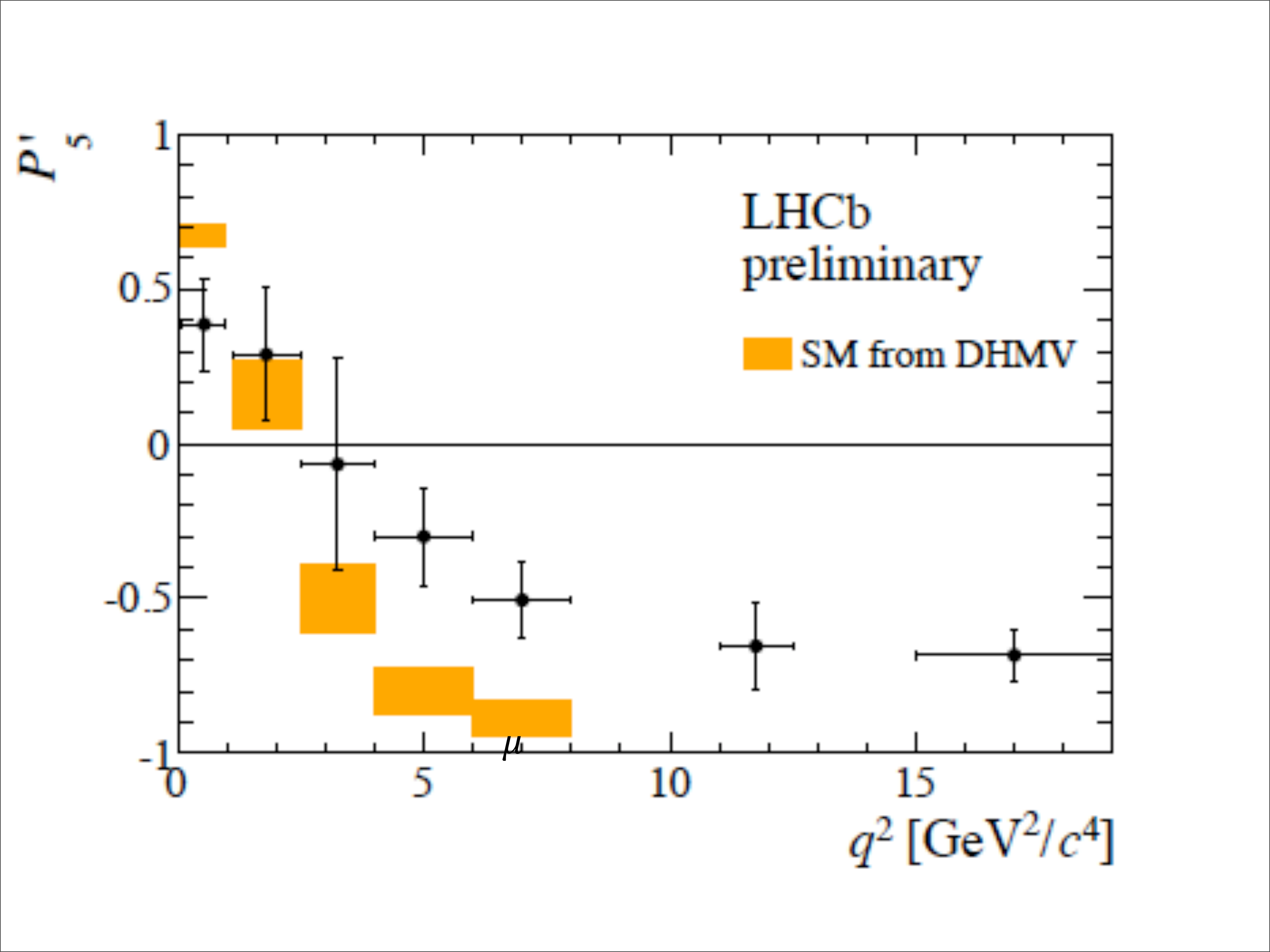}
\hfill
\includegraphics[angle=0,width=0.49\linewidth]{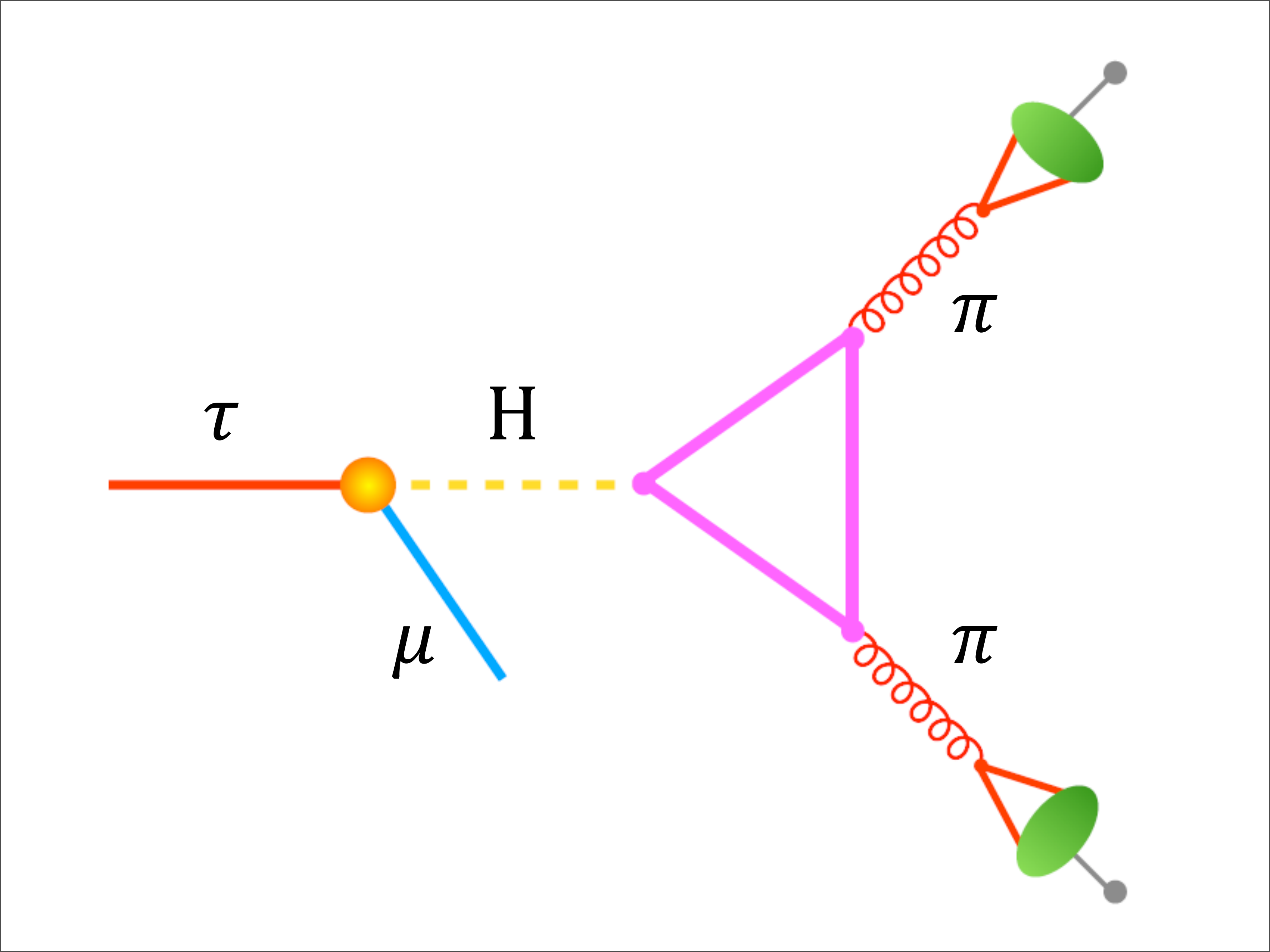}
\caption[]{Left plot: LHCb measurement of $P_5^\prime$ compared to SM predictions as a function of the momentum transfer. Plot taken from ref.~\cite{Aaij:2013cda}.
  Right plot: representation of the decay $\tau \to \mu H (\to \pi
  \pi) $.  }\label{fig:flav}
\end{figure}

Sebastian Descotes-Genon focused on the $B \to K^* \mu^+\mu^-$
anomaly, shown in Fig.~\ref{fig:flav} (left panel).~\cite{descotes} He
explained how different kinematic regimes at low or high momentum
transfer $q^2$ imply a high or low recoil of $K^*$ and hence NLO QCD
factorisation or heavy quark effective theory become
the appropriate tools to employ in the two regimes, respectively. He
also explained how observables like $P_5'$ are constructed in such a
way as to cancel large uncertainties from soft form factors and that residual effects
of power corrections are estimated
to be about 10\%.  
The conclusion of the study is that the LHCb measurement supports
$C_9^{\rm NP} \sim -1$, but there is room for NP also in other Wilson
coefficients. Discussions about the accuracy of the theory predictions
and the interpretation of data in terms of new physics are still
ongoing.

Andreas Crivellin pointed out that all four LHCb anomalies could be
explained in two models with gauged $L_\mu - L_\tau$: either a
two-Higgs doublet-model (2HDM) with vector-like quarks or a
three-Higgs doublet-model (3HDM) with gauged flavour dependent B-L
charges.~\cite{crivellin} Such a model predicts also a non-vanishing
$\tau \to 3\mu $ decay. Since the model has a point-like $H\mu\tau$
vertex, a question was raised whether this is consistent with current
limits on $\tau \to \mu \pi \pi$, which involves a Higgs exchange and
the decay of the Higgs to two gluons through a top loop, illustrated
in the right panel of Fig.~\ref{fig:flav}. After a quick calculation
Crivellin and Grinstein established that the model is still allowed.~\footnote{Greg Landsberg however pointed out that if you have
  four anomalies, and a model that explains all four of them, the
  model is wrong as at least one anomaly will surely go away.}
This example illustrated how difficult it is today to design new models that explain possible anomalies but are not yet excluded by precision data. 
Still looking at extended Higgs sectors, Eibun Senaha presented a
scale-invariant 2HDM with Coleman-Weinberg
(CW) symmetry breaking, rather than spontaneous symmetry breaking as
in the SM Higgs mechanism.~\cite{Senaha} The model predicts
deviations, for instance in the $h\to \gamma \gamma$ decay and in the $hhh$
coupling that are potentially detectable in future experiments.
Another 2HDM, a Branco-Grimus-Lavoura (BGL) model, with naturally suppressed FCNC
as a result of a symmetry of the Lagrangian 
was discussed by Gustavo Castelo-Branco.~\cite{castelo}
In BGL models the entire flavour structure is controlled by the CKM
matrix.  2HDM are an excellent framework to study NP in the scalar
sector. For instance, it is not clear yet whether the discovered Higgs
has small FCNCs. Hence, Branco pointed out that it is very important to
search for flavour violation in the scalar sector.  He stressed that
theorists have been wrong many times, it is therefore important to
look without prejudice, in particular to look also for
interactions or deviations that are not predicted in more common NP
scenarios.

A last theory talk in the flavour session was given by Cai-Dian Lu,
who pointed out that when the final state consists of two vector
particles, an angular study of the vector's decay products provides
an insight into the spin structure of flavour-changing
interactions.~\cite{lu} He stressed that, in particular, the meson
decays $B_{u,d,c,s} \to VV$ have a very rich phenomenology. Describing
them is however difficult and many theoretical approaches exist, but
most of them can not fully explain all measurements. He used a
perturbative QCD approach, and stressed the important role of
annihilation-type diagrams. Lu then presented many predictions
and comparisons to data. Some looked very successful, others
less so, however no theory uncertainties were given, so that it was
difficult to draw conclusions.

\section{QCD}

Almost all talks in the QCD sessions were theoretical.  The day was split
into a more perturbative morning focusing on calculations of
higher-order corrections, and a more non-perturbative afternoon,
mostly involving new theoretical developments.

\subsection{Perturbative QCD} 

First of all, it is important to remind ourselves why it is important to push the perturbative accuracy to higher orders. We have by now seen amazing results from  Run I at the LHC, including competitive measurements of SM parameters (even of the strong coupling constant $\alpha_s$), precision Higgs physics, jets spectra up to several TeVs, constraints on anomalous couplings, on NP models, on dark matter (DM) candidates and more,  as summarised by Tom Le Compte.~\cite{lecompte} Even better results are expected at Run II. An  optimal use of the machine can be achieved when the experimental (statistical and systematic) and theoretical uncertainties  are comparable. Currently, the use and interpretation of some cross sections is already limited by large theory uncertainties, mostly estimated via a variation of renormalisation and factorisation scales. Hence, it is mandatory to push the perturbative accuracy even further. 
Pier Paolo Mastrolia presented a beautiful review of modern methods for higher-order calculations.~\cite{mastrolia} He stressed the richness and power of factorisation, which is at the core of unitarity approaches and presented an interesting analogy between quantum mechanics and Feynman integrals. He then discussed a unitarity formalism, the Magnus Expansion, for multi-loop master integrals. Future directions include the treatment of multi-loop diagrams with internal masses and more legs and an automated analytic treatment of the one-loop case. Analytic results are in fact typically superior to numerical ones, since they are faster and numerically more stable. 

\begin{figure}[t]
\includegraphics[angle=0,width=0.49\linewidth]{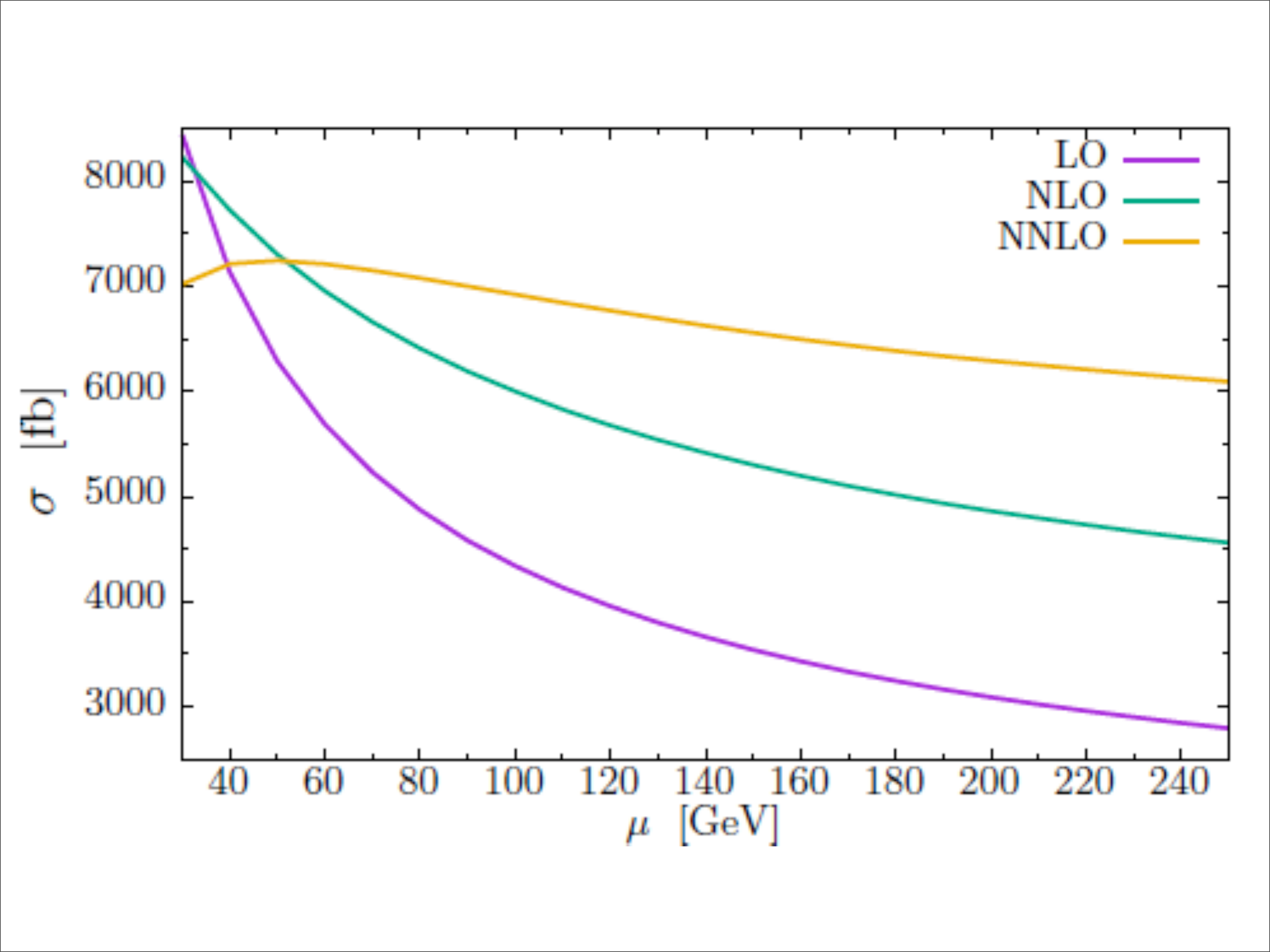}
\hfill
\includegraphics[angle=0,width=0.49\linewidth]{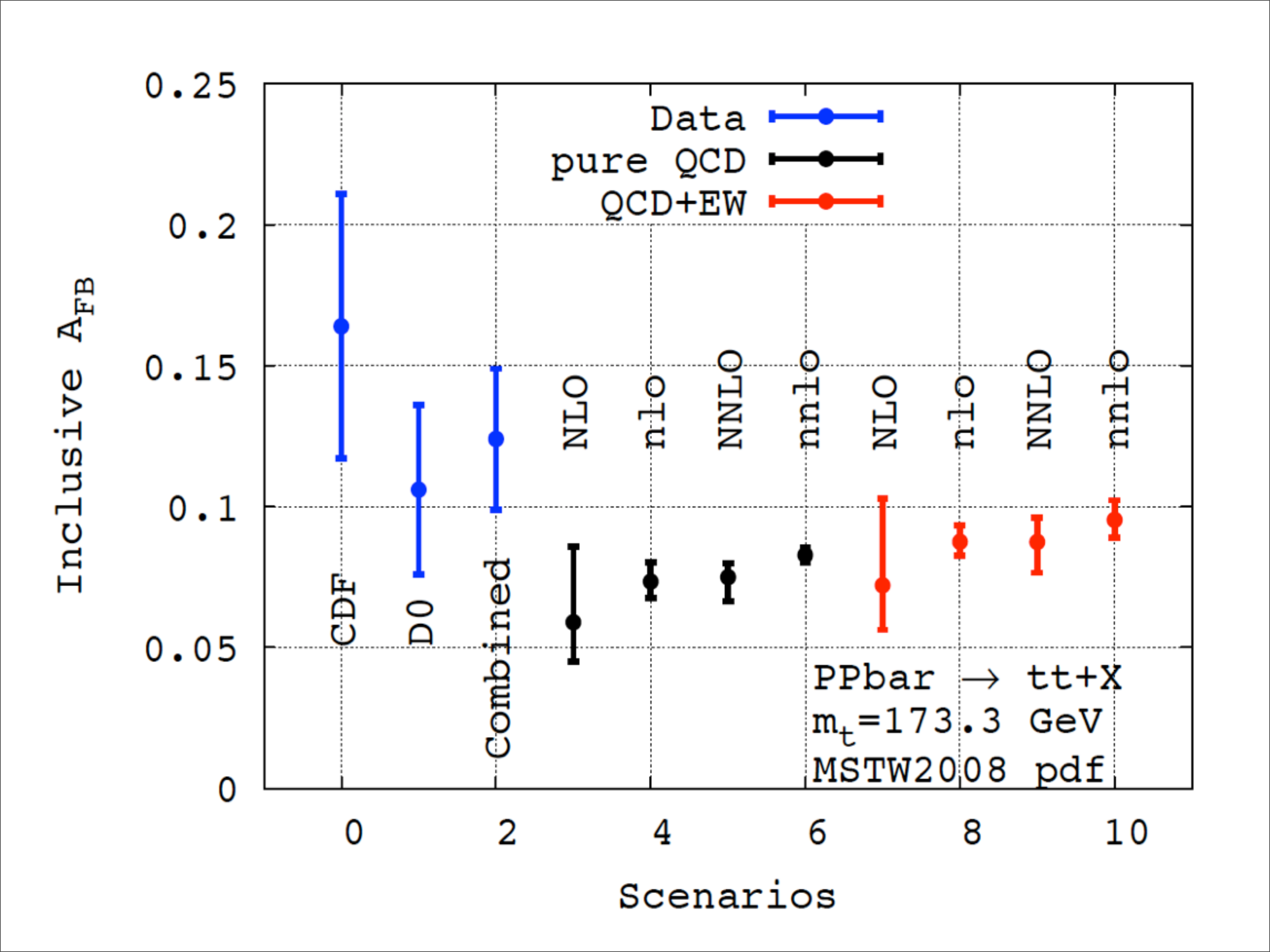}
\caption[]{Left plot: the Higgs plus one jet cross section at the 8
  TeV LHC at various orders in perturbation theory as a function of
  the renormalisation and factorisation scale, both set to $\mu$. Jets
  are defined using the anti-$k_t$ algorithm with $R=0.5$ and $p_{\rm t,
    cut} >$ 30 GeV.  Right plot: data for the the forward-backward
  asymmetry from D0 and CDF (blue) compared to various theory
  predictions including only QCD corrections (black) or including both
  QCD and EW corrections (red).  }\label{fig:pert}
\end{figure}

Jonas Lindert showed novel one-loop methods at work: he presented the
QCD and electroweak (EW) corrections to onshell $W$ production in association with 1, 2, or 3 jets obtained using {\tt OPENLOOPS}, {\tt MUNICH} and {\tt
  SHERPA}.~\cite{lindert} While NLO EW corrections might seem just a
trivial extension of the QCD case, they are in fact technically much
more complicated then just NLO QCD and involve a lot of subtleties.
The phenomenological results for $W$ plus multi-jet production that he
presented are very rich, in particular one can observe a non-trivial
dependence on the jet multiplicity. $W$ plus multi-jet events play a
key role for tests of the SM and for beyond SM (BSM)  searches based on signatures
with jets, a lepton and missing transverse energy (MET). One of the
main outcomes of the results presented by Jonas is that EW corrections are
important in the TeV region (where all order Sudakov effects should
also be included). For the future, the plan is to include vector
bosons decays, parton shower corrections and to extend  multi-jet
merging to NLO QCD+EW.

Beyond NLO corrections, Fabrizio Caola presented a brief review of the
status of NNLO, including the motivation to push the perturbative
accuracy to this order, the different methods used, the processes
known or almost known (Higgs, Drell-Yan, $t \bar t$, single top, $V$+1
jet, dijets, dibosons, $H$+1 jet).~\cite{caola} The main message from his talk
is very positive, i.e. that the NNLO technology is now ready to cope
with LHC demands. However, because the phenomenological environment is so rich,
it will not be enough here to provide numbers for cross sections or
few distributions, and there will be a lot of pressure on the authors
to release codes as soon as possible. He also showed results for Higgs
plus one jet production at NNLO, which at the time of the conference
were new and still unpublished. The NNLO corrections turn out to be
sizable (of the order of 20\% if the Higgs mass is taken as a central
renormalisation and factorisation scale), and reduce the scale-dependence of the cross section, as can be seen from
Fig.~\ref{fig:pert} (left). This was expected given the large
corrections in the case of inclusive Higgs production, and emphasizes
the importance of including NNLO corrections.

Another important NNLO calculation was presented by Michael Czakon,
who discussed the long-standing tension between SM predictions and
Tevatron measurements of the forward-backward (FB) asymmetry, for
which plenty of tentative BSM explanations have been given in the
past.~\cite{czakon} Recently, the SM theory prediction for the
asymmetry has been upgraded to include full NNLO corrections. A
limitation of the calculation is that the top quarks are stable, i.e. the result is fully inclusive in the top decay. One ambiguity
that arises is due to the fact that the FB asymmetry is the ratio of
the asymmetric over the symmetric cross section, hence one can choose
to expand the ratio in powers of the coupling constant, or not. These two approximations are denoted by
nnlo and NNLO in the right panel of Fig.~\ref{fig:pert},
respectively. Furthermore, there is an ambiguity in how to combine EW
and QCD corrections, either in an additive or in a multiplicative way.
Fig.~\ref{fig:pert} (right) shows that there is now perfect agreement
between the D0 result and NNLO theory, while the NNLO result is just $1.5\,\sigma$
below the CDF measurement. The fact that the discrepancy in the
asymmetry is not present any longer is mostly due to the updated
measurement of D0, however the NNLO calculation was very important to
confirm the robustness of the SM prediction. The calculation of the
NNLO top cross section was presented two years ago, it took then a
long time to provide predictions for the Tevatron asymmetry. One might
then wonder whether it is realistic to expect distributions for the LHC
in a reasonable amount of time, especially given the tension in the transverse momentum spectrum of boosted tops between next-to-leading-order (or LO matrix element plus parton shower generators) and ATLAS data.\cite{ATLAS:2014daa}  Michael Czakon is now working on a
completely new software based on a four dimensional subtraction
scheme, which will be several orders of magnitudes faster than the
first NNLO code for top-pair production.

\begin{figure}[t]
\includegraphics[angle=0,width=0.49\linewidth]{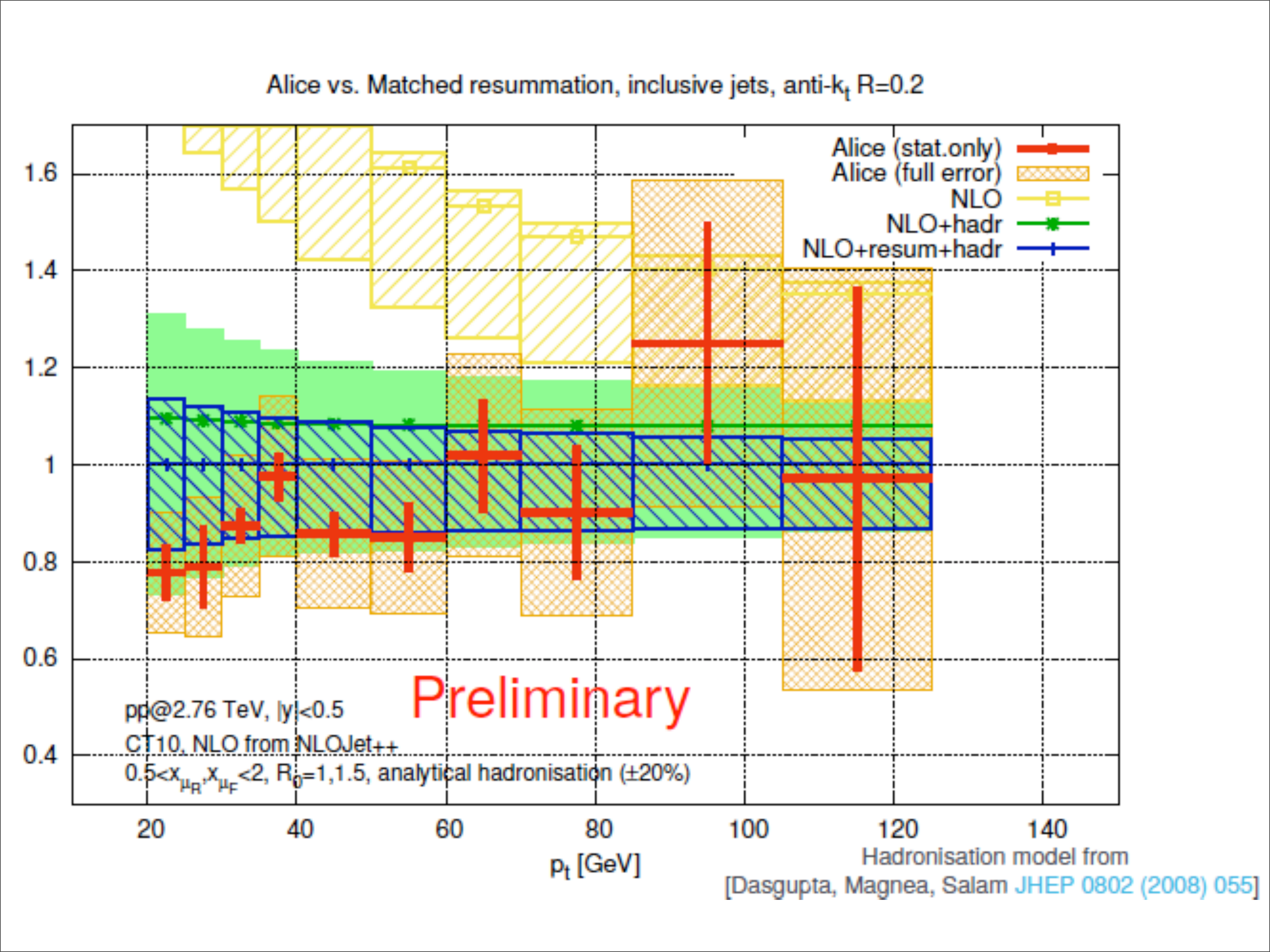}
\hfill
\includegraphics[angle=0,width=0.49\linewidth]{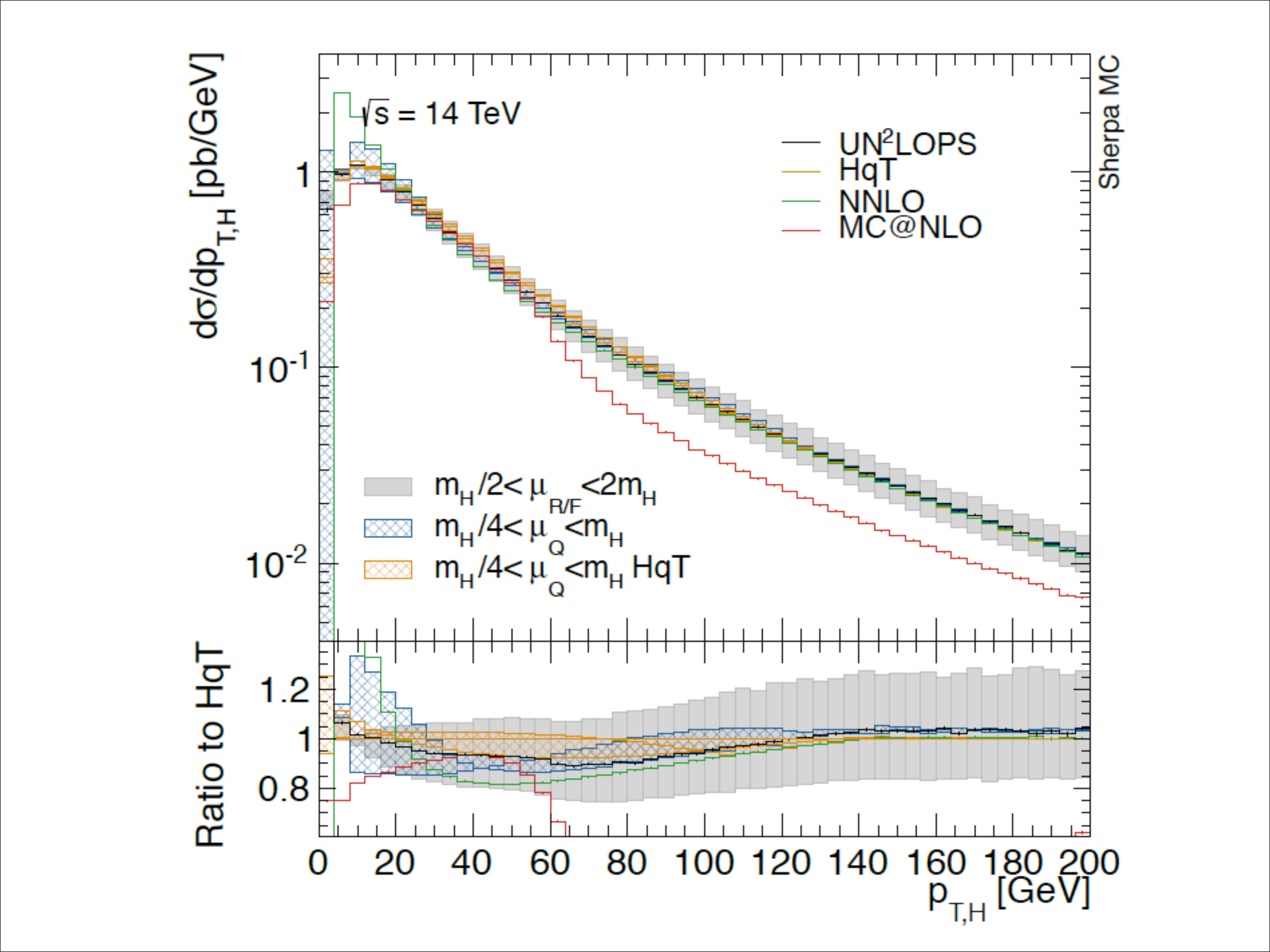}
\caption[]{Left plot: inclusive jet distribution measured at ALICE
  (red points and orange bands), compared to pure NLO with
  hadronisation corrections (green band) and a calculation that
  includes also $\ln(R)$ terms to all orders (blue band). Jets are defined using the
anti-$k_t$ algorithm with $R=0.2$. A reduction of
  theoretical uncertainty can be observed.  
  Right plot: the Higgs transverse momentum spectrum using the
  UN$^2$LOPS method, compared to other calculations.  
    }\label{fig:pert3}
\end{figure}

While fixed-order calculations are very valuable in general, they are
known to fail in particular regions of phase space. In these regions,
either parton showers or analytic resummations can be employed.
In Higgs studies involving a jet-veto, it was suggested some time ago
that there are potentially large logarithms related to the use of
small jet radii.  In fact, when the jet radius is small, the effect of
emissions outside the jet, that reduce the energy, can become
important. Because of the phase space constraints, this effect scales
as $\ln(R)$.
Frederic Dreyer explained that in a number recent jet-studies smaller
jet-radii $R$ are being used (e.g. $R=0.2$ in heavy-ions to mitigate
pile-up, or in jet-substructure studies).~\cite{dreyer} Frederic
showed that a resummation of leading $\ln(R)$ terms has been recently
carried out using an evolution equation for the quark generating
function.  Fig.~\ref{fig:pert3} (left) illustrates the reduced scale
dependence that can be achieved once the $\ln(R)$ resummation is
performed for inclusive jet production. The plot also shows a comparison
with experimental measurements from ALICE. There is good agreement
within the currently large experimental errors. It is clear that this
work will be even more relevant for future analyses, when more precise
data will be available and the use of smaller jet radii will become
more widespread in order to reduce increasing pileup contamination and
to study more highly collimated jets.

Still in the spirit of improving fixed-order calculations, Stefan Prestel discussed going beyond NNLO by merging NNLO and parton
showers.~\cite{prestel} This is important to have the best possible
perturbative prediction and the fully exclusive description (i.e. the
best of both worlds).  NNLO was recently merged to a parton shower in
the UNNLOPS approach for Drell Yan and Higgs production. Results for
the Higgs transverse momentum distribution are shown in
Fig.~\ref{fig:pert3} (right). Currently, within this approach, the
zero-$p_t$ bin is problematic, since the virtual correction is not spread
by the parton shower, but it sits all at zero transverse momentum. In
future, it would be desirable to extend the NNLOPS description to more
complicated processes, however such a task is not trivial no matter
which NNLOPS approach one considers.

\begin{figure}[t]
\includegraphics[angle=0,width=0.49\linewidth]{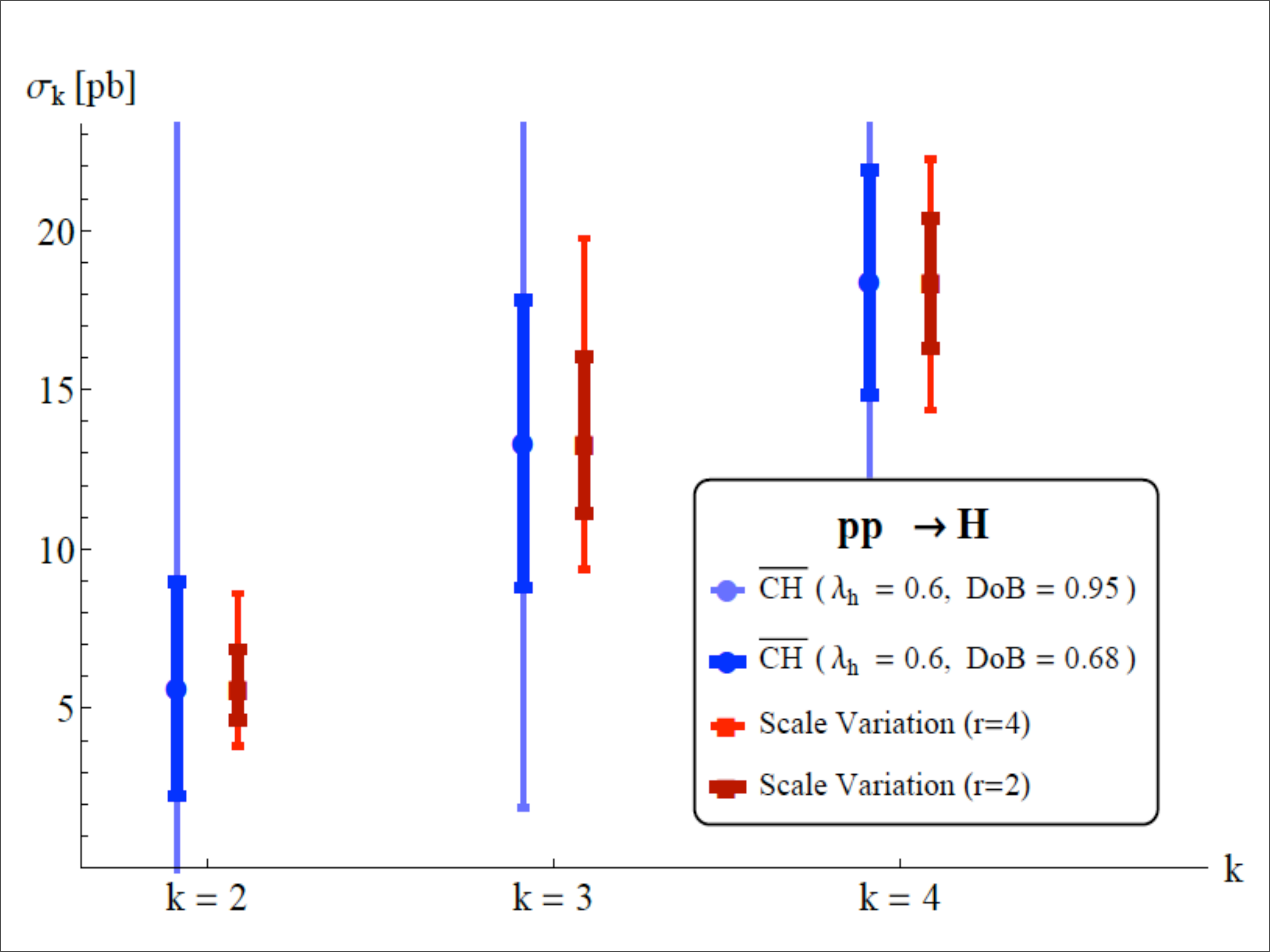}
\hfill
\includegraphics[angle=0,width=0.49\linewidth]{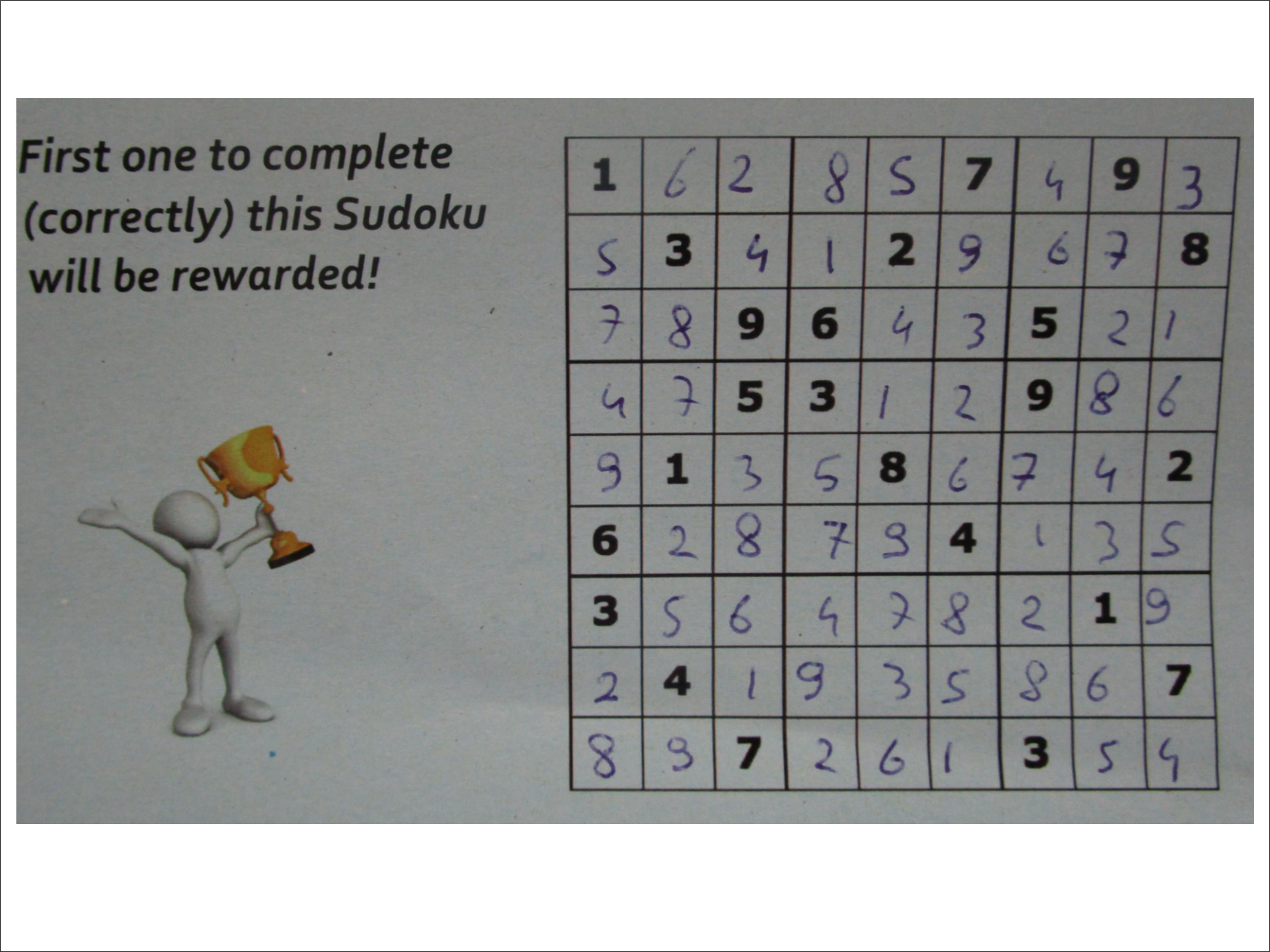}
\caption[]{
  Left plot: a
  comparison between two methods to estimate the theoretical
  uncertainties, scale variation (red) and the $\overline{\rm CH}$
  approach.
  Right plot: a (non-standard) application
  of random matrix theory.  }\label{fig:pert2}
\end{figure}

While the calculation of higher-order terms in the perturbative
expansion is obviously very useful to reduce the theoretical
uncertainty, it is also important to have a solid procedure to estimate
this residual uncertainty. This is obviously difficult, as
the knowledge of the next term in the expansion would be required to
provide a very reliable estimate of the theoretical uncertainty. 
A widely adopted procedure to estimate this uncertainty consists in varying the 
renormalisation and factorisation scales around a central value, which is chosen to reflect the hardness of the
hard process.
Emanuele Bagnaschi discussed how this scale-variation procedure has
severe limitations in estimating the true theory
uncertainty.~\cite{bagnaschi} On top of this, the theory uncertainty
has no statistical meaning, so it cannot be combined "properly" with
experimental statistical uncertainties. In 2011 Cacciari-Houdeau (CH) proposed a
Bayesian approach to estimate missing higher orders.  Recently the CH
method was modified to use a variable expansion parameter
($\overline{\rm CH}$). Furthermore, a first comprehensive set of more then 30 
observables was used to compare the $\overline{\rm
  CH}$ method to standard scale variation, including, for the first
time, hadronic observables.
Fig.~\ref{fig:pert2} (left) shows a comparison between the scale
variation uncertainties and the $\overline{\rm CH}$ uncertainties for
Higgs production. $k$ denotes the order at which the calculation is
performed. The scales have been varied by a factor 2 or a factor 4,
while for the $\overline{\rm CH}$ approach, the 68\% and 95\%
confidence intervals are shown. The general conclusion from this and
other plots is that in many cases, scale variation appears to do a good job in
estimating the size of the uncertainty. Furthermore, in most cases where the procedure fails, we believe we 
understand the reason. On the other hand there is value in having a
quantitative, statistical meaning to any statement referring to theory
uncertainties. In future, one can expect many valuable comparisons
between the $\overline{\rm CH}$ approach and standard scale variation
using new NNLO calculations that are becoming available.

Another source of theoretical uncertainty, beyond missing higher
orders comes from our limited knowledge of the strong coupling
constant.
David D'Enterria pointed out that in fact $\alpha_s$ is to date the
least precise known of all couplings (known to about (0.5-1)\%), and
this impacts all LHC cross sections.~\cite{denterria} Furthermore, a
very precise knowledge of the strong coupling is a key for SM
precision fits and is relevant in BSM studies (e.g. for coupling
unification at GUT). The current world average is $\alpha_s=0.1185\pm
0.0006$. David presented new fits of $\alpha_s$ using the jet
fragmentation function in $e^+e^-$ and DIS data using an approximate
NNLO calculation matched to NNLL. These fits give a value of the
coupling of $\alpha_s=0.1205\pm0.0010$. In future the plan is to
extend the calculation to full NNLO+NNLL. Still, these fits are by
far not trivial, as they require a careful treatment of the
correlation between data and of heavy-quark thresholds.

\subsection{Less perturbative QCD} 
In-between fixed-order expansions and non-perturbative regions, there are regions
of phase space where calculations are needed that resum large classes of corrections 
to all orders in the coupling constant, typically those accompanied by large logarithms.
Many of these resummed calculations rely on the formulation of a
factorisation theorem. Such a factorisation, while not necessary,
turns out to be very useful in many cases, as it allows one to split
the calculations into various elements that are simpler to calculate and that can
be computed in one context and used in a different one. 
Mark Harley presented a  clear introduction to some of these ingredients, i.e. Wilson lines,
soft functions, (cusp) anomalous dimensions and webs.~\cite{harley}
The aim of his work is a better description of universal soft
singularities. Webs in this respect are very useful: they organise diagrammatic
contributions to the exponent of the soft function, and one can show that
all such contributions appear with connected colour factors. Subtracted webs 
are what remains in the exponent after the removal of multiple UV poles. Finally,
multiple gluon exchange webs (MGEW) are subtracted webs with no gluon
self-interaction (i.e. only dipole-like exchanges). It has been conjectured, and 
confirmed by explicit calculations in specific cases, that the integrand of MGEWs 
contains no polylogarithms, only logarithmic functions, each dependent on a 
single cusp angle. It is still an open question whether this always holds for MGEWs 
and it is not clear why does this happen. Furthermore, work is under way to extend 
current techniques to compute more general webs: in particular, results for completely 
connected diagrams at three loops were recently announced at the Radcor-Loopfest 
conference.~\cite{gardi}

Regarding the treatment of radiation from the initial state partons, all higher-order calculations mentioned so far rely now on collinear factorisation, which however does not work well when one incoming parton carries a very low momentum fraction. 
Sebastian Sapeta discussed an improved transverse momentum dependent
factorisation for forward dijet production in dense (small $x$)-dilute
(large $x$) hadronic collisions.~\cite{sapeta} This is the only existing
approach which is valid in all regions of the transverse momentum of the target, from
very high transverse momenta, where high-energy factorisation is
usually applied, to very low ones, where collinear factorisation
holds. Hence, this approach provides a robust framework for studies of
saturation domains with hard probes. The final aim is to gain a better
understanding of factorisation breaking and the nucleon structure.

While perturbative corrections can be calculated, non-perturbative
corrections are usually just modelled.  Sharka Todorova-Nova pointed
out that the Lund string fragmentation model has been very successful. 
It has been implemented in Pythia and, after tuning, it describes data well in general. Still, it has
limitations and some data is not well described by it. Hence, she
presented a study of quantum properties of three-dimensional helix-shaped QCD
strings.~\cite{todorova} The model is predictive after fixing two
parameters for the string. According to Sharka, in the near future it
will be possible to compare predictions from the model with upcoming
measurements.
Shi-Yuan Li pointed out that colour connections are the bridge between
parton and hadron systems.~\cite{Li} Four-quark systems ($ccbb$,
$bbbb$, etc.) have an intrinsic ambiguity in the colour wave-function,
which leads to different meson production. Li encouraged
phenomenological studies in $e^+e^-$ collisions to look and interpret
different meson production as evidence of certain colour connections.

Various approaches to non-perturbative dynamics use symmetries
and dualities to obtain results at strong coupling.  Miguel Costa
presented an AdS/QCD phenomenological model that matches well the
intercept and slope of Donnachie-Landshoff pomeron.~\cite{Costa} The
model is predictive, since everything in the model is fixed from soft
pomeron exchange. A careful analysis of data for deep inelastic scattering (DIS), deeply virtual Compton scattering (DVSC) and virtual meson production (VMP) is
hence interesting.  Andrew Koshelkin looked at multi-particle dynamics
and pion production using a flux tube (a compactification to two dimensions) and
showed a comparison to ALICE data for the transverse momentum
distributions in high-energy proton-proton
collisions.~\cite{Koshelkin} Giancarlo D'Ambrosio pointed out that
soft wall models in holographic QCD have correct Regge trajectories
but a wrong operator product expansion (OPE).~\cite{Ambrosio} Hence, he
presented a modified version of the dilaton potential that allows one
to comply OPE. OPE is recovered by adding a boundary term.  Low energy
chiral parameters, $F_\pi$ and $L_{10}$, are well described
analytically by the model in terms of Regge spacing and QCD
condensates.

We had a single lattice talk at the meeting on random matrix
theory. Chiral Random Matrix Theory is a powerful mathematical tool to
calculate eigenvalue correlations in the IR limit of QCD.  The way it
works is simply to replace the Hamiltonian with a Random Matrix with the
same global properties. Once this is done, one can compute observables
by averaging over ensembles. Here it is critical to identify what are
the universal quantities (i.e. those that are independent of the 
probability distribution). Savvas Zafeiropoulos considered explicitly
the case of $N_c = 2$ QCD and presented a study of the discretisation
for $D_5$ (i.e. the hermitian version of Wilson operator $D_W$) and
$D_W$ itself.~\cite{Zafeiropoulos} In the future he plans to study the
case of adjoint QCD, which has a different chiral symmetry breaking
mechanism.~\footnote{An application of random matrix theory, in a
  different context, is illustrated in Fig.~\ref{fig:pert2} (right
  panel).}

\section{New Phenomena}
\begin{figure}[t]
\includegraphics[angle=0,width=0.49\linewidth]{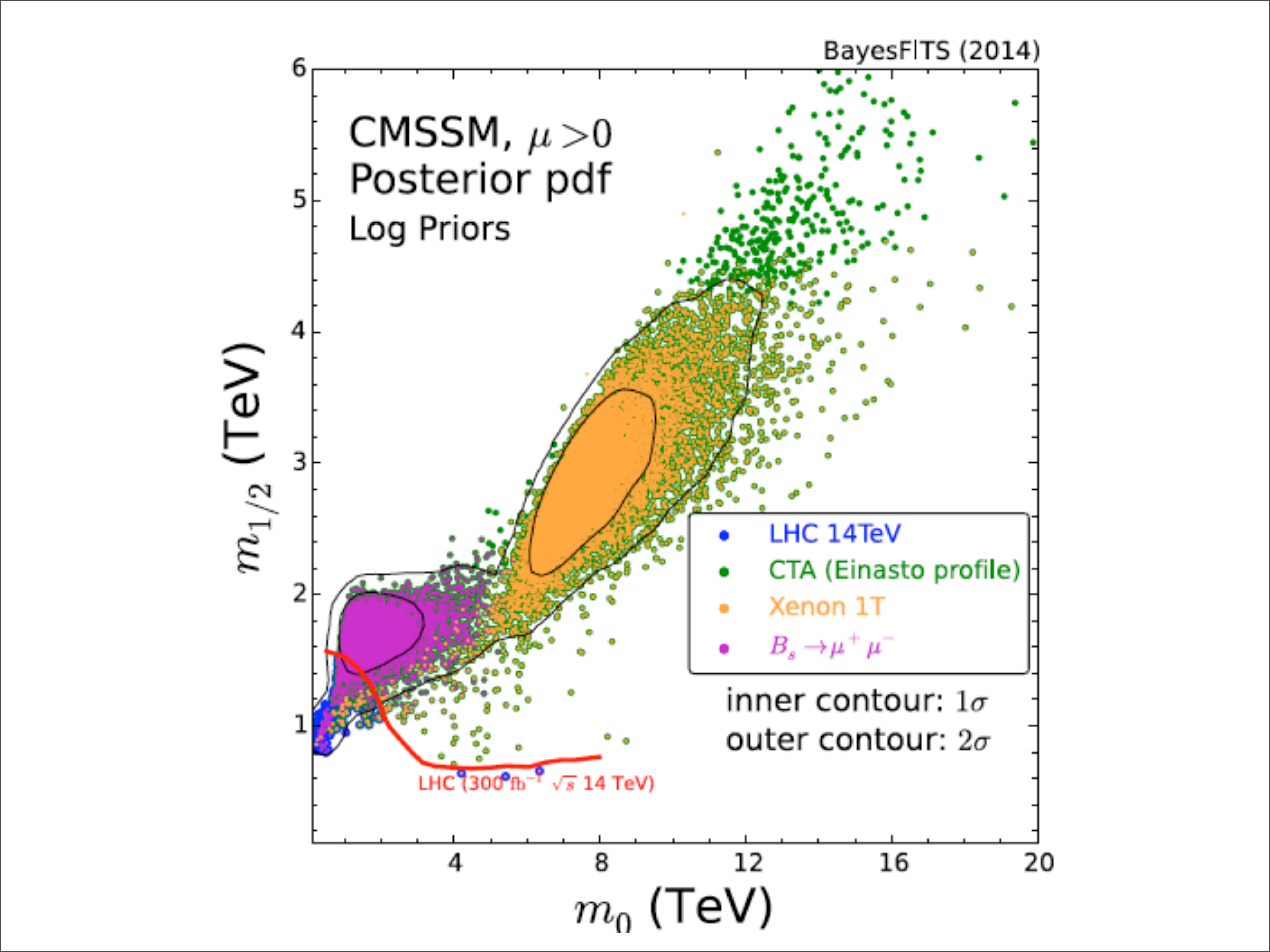}
\hfill
\includegraphics[angle=0,width=0.49\linewidth]{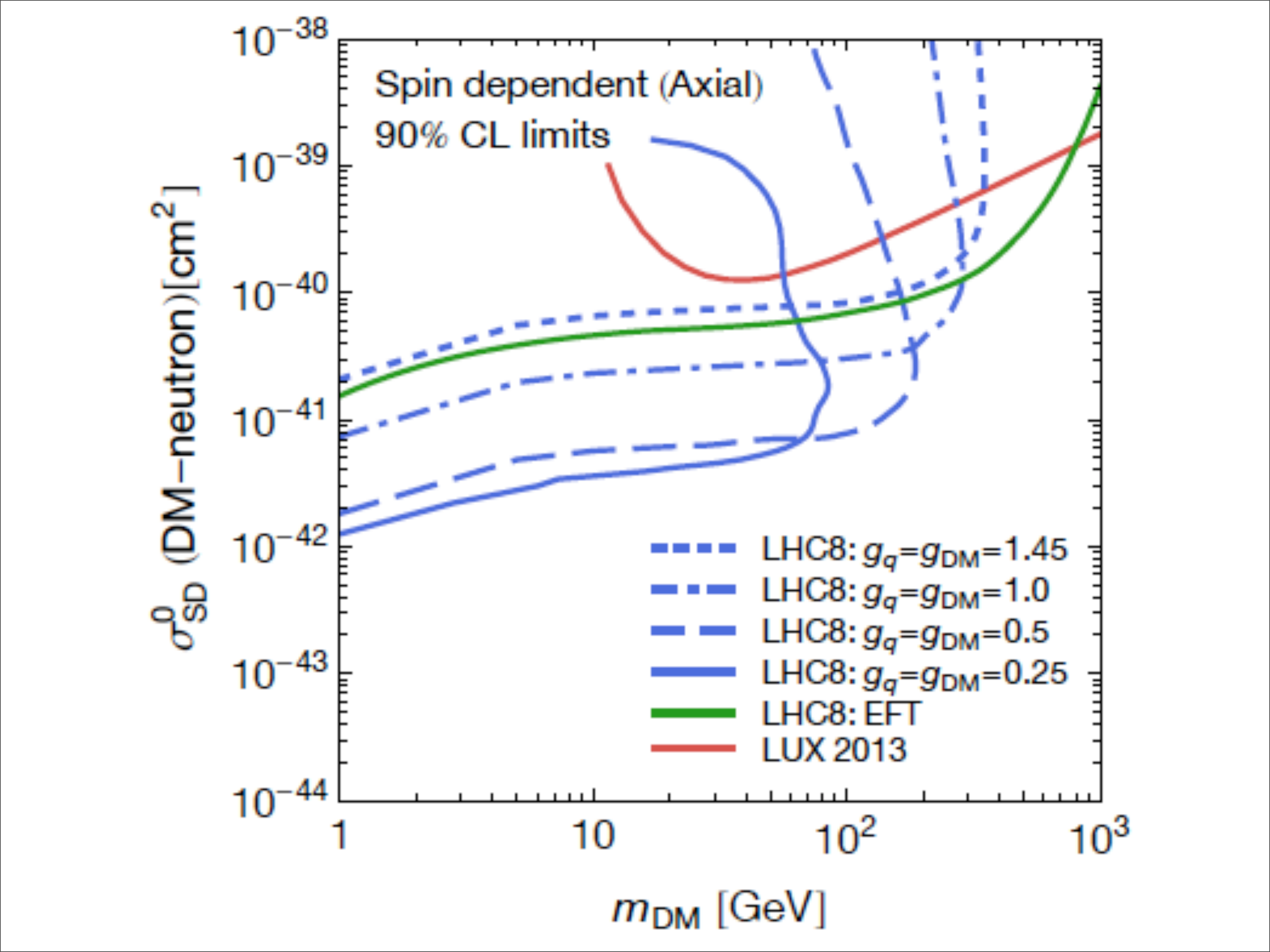}
\caption[]{Left plot: illustration of the range of the $m_{1/2}$ and
  $m_0$ plane that can be explored by various experiments.  Right
  plot: bounds for the spin-dependent cross section that can be
  obtained within an EFT approach (green), or using a simplified model
  with different choices of the coupling of the mediator to quarks and
  DM (blue). Bounds from LUX (red) are also shown.  Plot taken from
  ref. \cite{Malik:2014ggr}.  }\label{fig:NP}
\end{figure}

Unfortunately, no evidence for new phenomena has been
seen in Run I at the LHC.  Possibly the strongest motivation for
physics beyond the SM is the astrophysical evidence for DM in
galaxies and in cosmological observations.
Leszek Roszkowski pointed out that the measured value of the Higgs
mass of 125 GeV allows for rather heavy SUSY states, too heavy to be
produced at the LHC. DM searches provide then important
complementary bounds to collider searches.~\cite{Roszkowski} In
particular he emphasised the possible future role of the Cherenkov
Telescope Array (CTA) experiment, a next generation ground-based very
high energy gamma-ray instrument. Beside exploring the origin of
cosmic rays and their role in the Universe, the nature and variety of
particle acceleration around black holes, the CTA aims at searching
for the ultimate nature of matter and physics beyond the Standard
Model. For instance within the CMSSM, it will allow one to explore
mass ranges (for $m_0$ and $m_{1/2}$) that are out of reach at the
LHC, as can be seen from Fig.~\ref{fig:NP} (left). On the other hand,
the existing tension with $(g-2)_\mu$, if taken seriously, requires
light-ish SUSY particles, which should be within LHC reach.

Another possible future experiment that is complementary to the LHC is
SHIP (Search of Hidden Particles). Oleg Ruchayskiy pointed out that
we might not have detected new particles at the LHC either 
because they are  too heavy, or because they are light but very weakly
coupled.~\cite{Ruchayski} The first case, will be investigated by energy
frontier experiments, currently the LHC Run II.  The second option
can be explored by going to the so-called intensity frontier. This is
precisely what SHIP aims to do. The idea is just to take a highest
energy/intensity proton beam, dump it into a target, followed by the
closest, longest and widest possible and technically feasible decay
tunnel. Oleg showed that for instance a Neutrino minimal SM, which
addresses neutrino oscillations, DM, baryon asymmetry, and inflation,
an be explored with SHIP. Similarly, SHIP can explore other models
that involve very weakly interacting long lived particles including
Heavy Neutral Leptons, right-handed partners of the active neutrinos,
light supersymmetric particles (sgoldstinos, etc.), scalar, axion and
vector portals to a hidden sector.
   
The complementarity between direct and indirect DM detection experiments
and the LHC was also stressed by Greg Landsberg.~\cite{Landsberg} In
fact, freeze-out, direct detection and collider production can be all
represented using the same diagram and crossing the direction of time.
DM is typically searched for at the LHC through so-called
mono-X searches, i.e. the production of one (or more) SM particles
or jets accompanied by a large MET that is
attributed to DM particles escaping detection.
Often DM searches use an EFT where the mediator has
been integrated out. Greg pointed out that since the mediator is
integrated out, the dynamics might not be properly described, and
hence the interpretation of EFT results becomes problematic.
Following the strategy used in SUSY searches, Greg suggested to use
so-called simplified models. Here one identifies simple models and
works out the signatures. The simplification limits the number of
arbitrary parameters, still providing a robust benchmark.~\footnote{It
  is important to keep in mind that while simplified models are
  practical tools, they can exhaust their value as a benchmark at some
  point, as, I believe, is the case for the CMSSM now.} The simplified
model for DM searches used by Greg uses an $s$-channel spin-1 mediator,
that interacts to DM and fermions. This is a four-parameter
model. Extensions for instance to include the case of $t$-channel
mediators, or spin-0 interactions are also possible.
Fig.~\ref{fig:NP} (right) shows a comparison of EFT bounds
(green) to the above simplified model assuming the spin-1 mediator to 
have pure axial-vector couplings. It is evident that in some regions of parameter space EFT
results are too optimistic, in others they provide too loose bounds.

\section{Heavy Ions}
\begin{figure}[t]
\includegraphics[angle=0,width=0.49\linewidth]{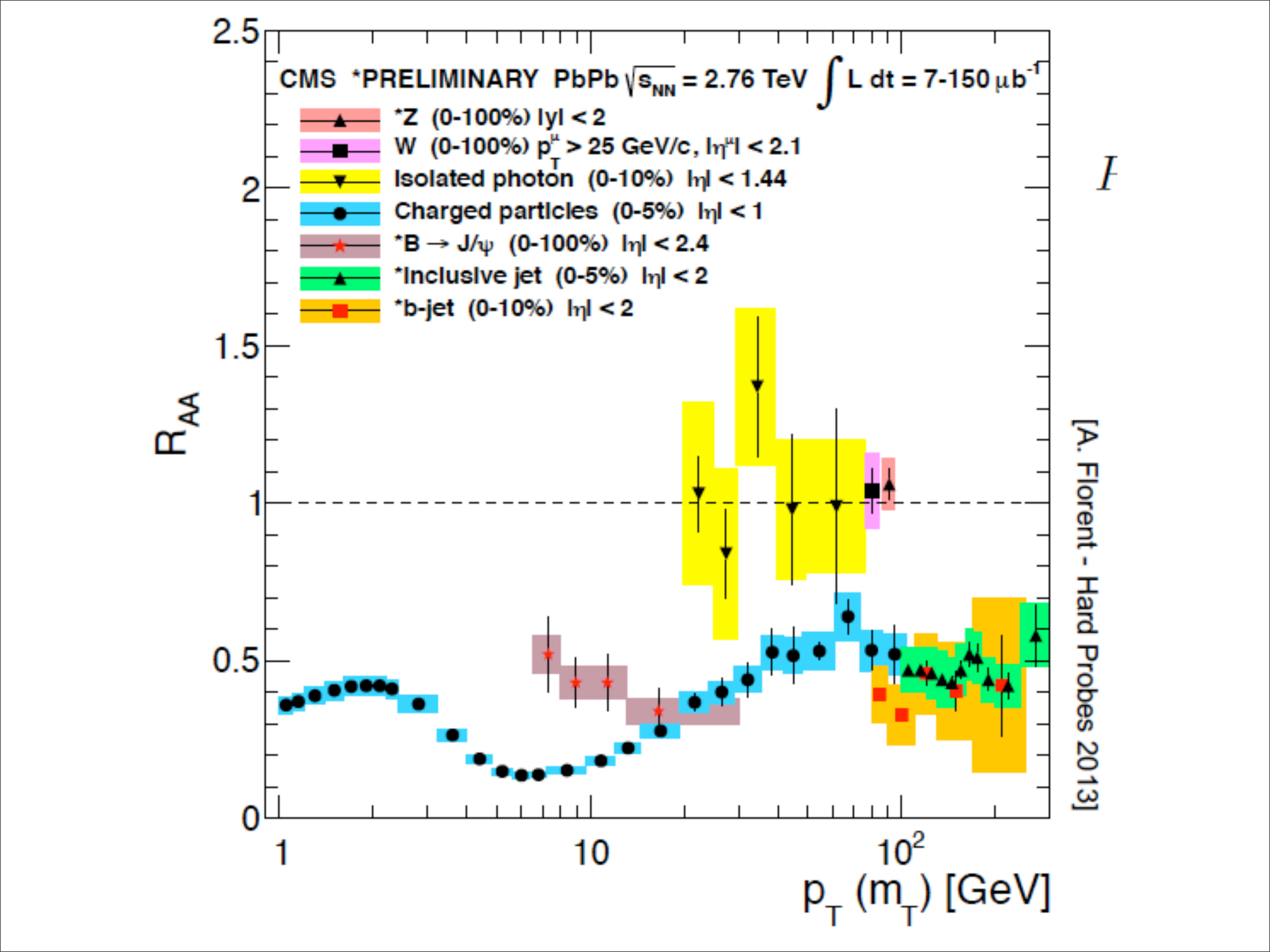}
\hfill
\includegraphics[angle=0,width=0.49\linewidth]{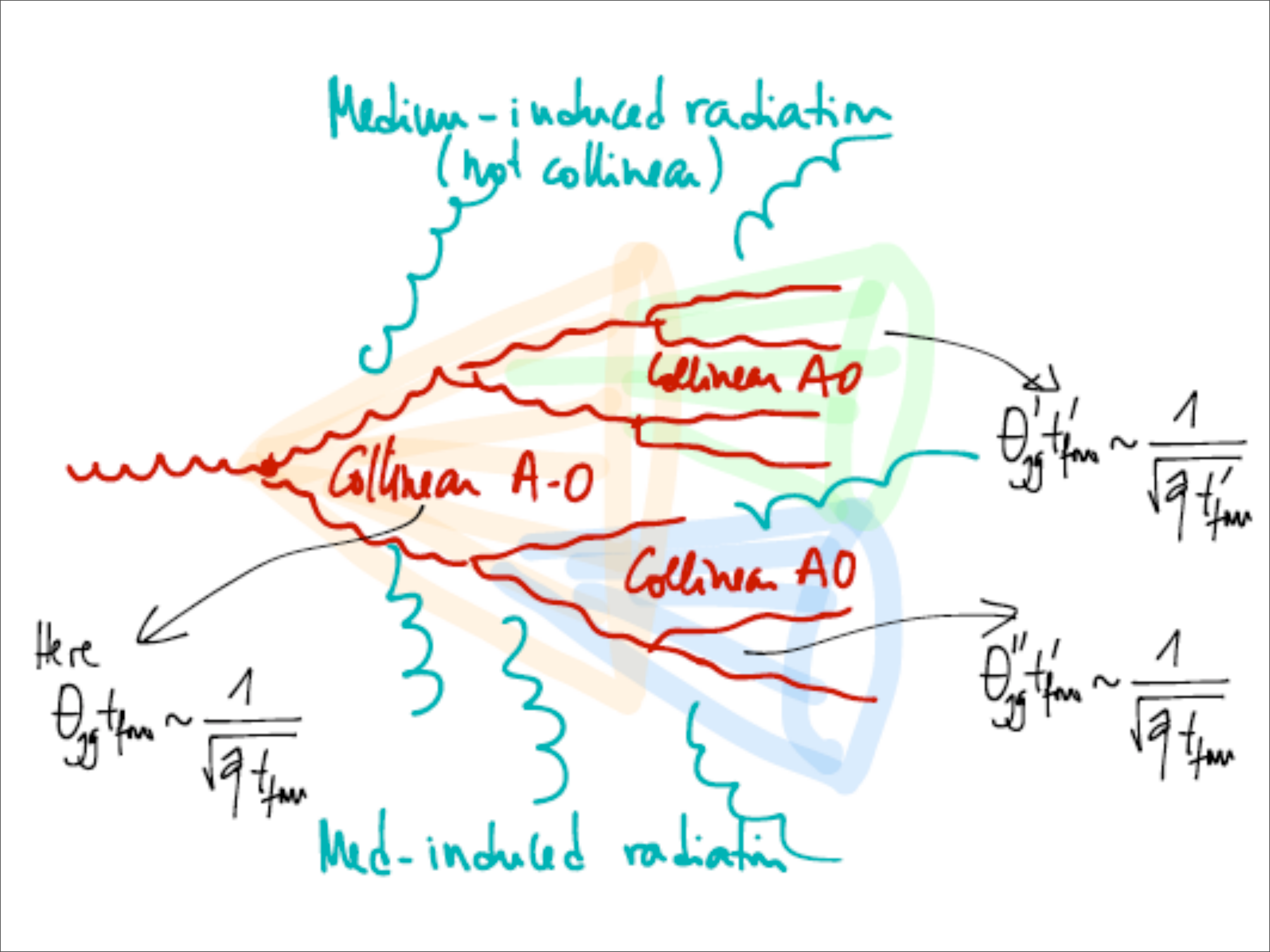}
\caption[]{Left plot: the nuclear modification ratio as a function of
  the transverse momentum.  Right plot: a new picture of jet-quenching
  that involves two components, a vacuum-like, angular-ordered
  component (red) and a medium-induced, not collinear radiation
  (blue).  }\label{fig:HI}
\end{figure}

Our Friday started with a very comprehensive introductory talk on
heavy ions (HI) by Carlos Salgado.~\cite{Salgado} He pointed out that
behind a simple QCD Lagrangian, there are very rich emerging
phenomena, like asymptotic freedom, confinement, chiral symmetry
breaking, mass generation, new phases of matter, and a very rich
hadron spectrum. Some of the questions raised by observations
can be further studied with heavy-ion collisions. For instance
one can study the structure of the hadrons and nuclei at high energy,
one can try to understand if the created medium is thermalised and
what are the properties of the produced medium. In this context, for a
long time proton-nucleon collisions (pA) were considered a benchmark
point needed to subtract the background from nucleon-nucleon (AA)
collisions. However pA collisions seems to have taken up an unexpected
role since results for pA collisions turn out to have some features
similar to those for AA collisions, in particular concerning the
collective, hydro-dynamical behaviour. This raises the important
question of whether such a small system can also thermalise.
One of the standard probes of a hydro-dynamical behaviour is the
so-called elliptic flow, i.e. the flow due to the fact that there is
more momentum in the plane of the collision, compared to the
transverse direction (a simple consequence of having a higher pressure
gradient in the plane).
Data are based on measuring two or more particle
correlations. Recently, there has been a lot of theoretical work in
trying to add fluctuations in initial conditions and in including
viscosity corrections.
Another standard tool is measuring jet quenching in medium. The idea
here is simply that a medium suppresses the propagation of coloured
particles, compared to the free propagation, hence this results in jet
suppression.
The simplest observable of jets in nuclear collisions is the
measurement of the one-particle inclusive production at high
transverse momentum. The effect of the surrounding matter can then be
identified by the suppression of the signal, with respect to the
proton-proton collisions, due to energy loss. A standard probe of
medium effects uses the nuclear modification ratio, defined as
\begin{equation}
R_{AA} \equiv \frac{d\sigma^{AA}/dydp_T}{N_{Coll} d\sigma^{pp}/dydp_T}\,,
\end{equation} 
where $N_{coll}$ is a normalization factor computed in the Glauber
model to allow the comparison with the proton-proton cross
section. The suppression of high-p$_{T}$ hadrons was one of the first,
and also one of the main, observations at RHIC.
Fig.~\ref{fig:HI} (left) shows the effect of the suppression at the
LHC, where the propagation of all particles interacting with the
medium are suppressed.  This can be seen from the fact that $R_{AA}$
is smaller than one, while for isolated photons and EW
bosons, that do not interact with the medium, $R_{AA}$ is compatible
with one. Carlos then presented a new picture of jet-quenching,
illustrated in Fig.~\ref{fig:HI} (right), where the parton shower is
composed of two overlapping components, which can be understood as a
reorganisation of the jet into multi-jets, with vacuum-like collinear
radiation, that effectively act as single emitters for the
medium-induced radiation.  It will be interesting to see how future
measurements compare to this new picture.
 
When studying correlations, Matt Luzum pointed out that it is
instructive to study two-particle correlation as a function of the
transverse momentum vector of the particles.~\cite{Luzum} In fact, the
hydro-dynamic behavior imposes constraints on the momentum structure of
two-particle correlation. One can define a full correlation matrix
\begin{equation}
  V_{n\Delta}(p_T^a,p_T^b) = \left \langle  \frac{1}{N^{a,b}_{\rm pairs}} \sum_{\rm pairs\{a,b\}} \cos n\Delta \Phi  \right \rangle\,,
\end{equation} 
where $N_{a,b}$ is the number of particles with momenta $p_T^a$ and
$p_T^b$ in a given event, $\sum_{\rm pairs\{a,b\}}$ is the summation
over all sets of these pairs, and $\Delta \Phi = \Phi^a-\Phi^b$ their
relative azimuthal angle.  This quantity can be used to define
\begin{equation}
  r_n \equiv \frac{V_{n\Delta}(p_T^a,p_T^b)}{\sqrt{V_{n\Delta}(p_T^a,p_T^a)V_{n\Delta}(p_T^b,p_T^b)}}\,.
\end{equation} 
Values of $r_n=1$ indicate no correlations, while $|r_n| > 1$ are a
sign of non-flow dynamics.  CMS and ALICE data for $r_2$ and $r_3$ was
also shown, however more data are needed, I believe, to draw solid
conclusions.
 
Alexey Boyarsky discussed the chiral magnetic effect.~\cite{Boyarsky}
This is experimentally searched for at RHIC through a charge asymmetry
of particles. It is usually widely discussed in the context of
quark-gluon plasma and heavy ions. This effect is related to the fact
that the SM plasma at finite densities of lepton and
baryon numbers becomes unstable and tends to develop large-scale
magnetic fields. The goal of Alexey's talk was to show that this
effect is more general, and has to do with relativistic plasma of
charged fermions (leptons and quarks). The conclusion is then that
this effect can be important in other contexts whenever  you have a relativistic
magnetised plasma (such as in the early universe, in neutron stars, in
astrophysical jets, and in quark-gluon plasma).
Elena Petreska discussed magnetic Wilson loops in the classical field
of high-energy HI collisions.~\cite{Petreska} In the abelian
case, the Wilson loop measures simply a flux, while in the non-abelian
case the Wilson loop obey area law for uncorrelated magnetic vortices,
which is found to hold for large enough areas.
  
Alexander Bylinkin discussed the origin of the thermal component in
transverse momentum spectra in high-energy hadronic
collisions.~\cite{Bylinkin} It is well-known that black holes radiate
thermal radiation with a temperature that is proportional to the
acceleration of gravity at the surface. Similarly, an observer moving
with acceleration $a$ detects a thermal radiation proportional to his
acceleration. This is usually referred to as Unruh radiation. In both
cases, the effect is due to the presence of an event horizon, for
instance, in the accelerated frame, part of the space-time is causally
disconnected from the accelerating observer. In the case of
high-energy collisions, confinement is proposed to produce the
effective event horizon for coloured particles. This results then in
thermal hadron production with  temperature of about 160
MeV. Bylinkin used these observations to present a two-component model
for hadro-production.  The two components are attributed to two
different mechanisms: hard radiation with a saturation scale, and a
thermal Unruh-like radiation. Bylinkin showed that there is  good
agreement between the available experimental data and the predictions
of the model for rapidity distributions, average transverse momentum
as a function of multiplicity, and transverse momentum spectra, which
have been notoriously difficult to describe with standard approaches.

\section{Looking ahead}
I am very much looking forward to coming Moriond meetings with lots of
exciting new experimental data. 
  While we all have high hopes, we also wonder what will happen
if, despite the tremendous experimental and theoretical efforts, we do
find any sign of new physics in Run II at the LHC. It is important to
remember that exploring the unknown is valuable in it's own
right. Surely it will not be wise to draw too quick conclusions, but
whatever happens, we will learn something by going to a new frontier
(Run II, HL-LHC, FCC,  ...).

\section*{Acknowledgments}

I would like to thank the organisers for the invitation to participate
to this splendid meeting and all participants for many stimulating
discussions and new insights. My work is supported by the ERC
consolidator grant HICCUP (614577). I would also like to express a special 
thanks to the Mainz Institute for Theoretical Physics (MITP) for its hospitality 
while completing this writeup.

\end{document}